\documentclass[a4paper]{article}
\usepackage{amssymb}

\usepackage{amsthm,amsmath,amssymb,amsfonts}

\newtheorem{The}{Theorem}

\newtheorem{Lemma}{Lemma}

\begin{document}

\title{Partially integrable nonlinear equations with one higher
symmetry}


\author{Alexander V. Mikhailov$^{+}\footnote{
On leave, Landau Institute for Theoretical Physics, Moscow, Russia}$, Vladimir S. Novikov$^{\& *}$ and Jing Ping Wang$ ^\& $\\
$+$ Applied Mathematics Department, University of Leeds, UK\\
$\&$ Institute of Mathematics and Statistics, University of Kent,
UK}

\maketitle


\begin{abstract}
In this paper we present a family of second order in time nonlinear
partial differential equations, which have only one higher symmetry.
These equations are not integrable, but have a solution depending on
one arbitrary function.
\end{abstract}

\section{Introduction}

For a long time there has been a general belief that "if a partial
differential equation or a system of differential equations has one
nontrivial symmetry, then it has infinitely many"
\cite{mr81i:35144}. Indeed, this statement is true in the case of
evolutionary equations, which right hand side is a homogeneous
differential polynomial \cite{mr99g:35058}. However, in 1991 Bakirov
proposed an example putting this conjecture in doubt \cite{ba91}. He
found that the system
\begin{equation}\label{Bak}
\{\begin{array}{l}u_t=u_{xxxx}+v^2\\
v_t=\frac{1}{5}v_{xxxx}\end{array}
\end{equation}
has a symmetry of order $6$ and he also verified using a computer
algebra software that it does not have other symmetries up to the
order 53. Recently, using $p$-adic analysis, it has been
 rigorously proven that system (\ref{Bak}) has
only one higher symmetry and therefore the above conjecture is not
valid \cite{mr99i:35005}. Furthermore, there exist infinitely many
systems of the same type with only one high symmetry. Even a
modified conjecture, stating that existence of $n$ higher symmetries
for $n$-component system of evolutionary equations implies
infinitely many \cite{fok87} is not valid either - there exists a
system, similar to (\ref{Bak}) of order 7, which has only two higher
symmetries \cite{mr2003j:37109}.

The Cauchy problem for systems of the form (\ref{Bak}) is always
solvable no matter how many symmetries the system has. One can solve
the second equation and substitute the solution into the first one.
The equation obtained  becomes a linear inhomogeneous equation,
which can be solved using standard techniques.

In this letter we present the following family of partial
differential equations $n\ge 2$
\begin{eqnarray}
\label{utt}
u_{tt}&=&u_{n,t}-\frac{u_t}{u}(u_n-u_t)+\frac{(u_n-u_t)^2}{u}-u\partial_x^n\left(\frac{u_n-u_t}{u}\right).
\end{eqnarray}
Here and further below we adopt the notation
$u_k\equiv\partial_x^k(u)$. The result of our study can be
formulated as the following

\begin{The}
For any $n\ge 2$ equation (\ref{utt}) possess a local higher
symmetry
\begin{equation}
\label{sym} u_{\tau}=\frac{1}{u}(u_n-u_t).
\end{equation}
This symmetry is the only local higher symmetry of equation
(\ref{utt}).
\end{The}

Equation (\ref{utt}) is not integrable by any known methods.
Nevertheless there is a family of exact solution depending on one
arbitrary function.

\section{Proof of the Theorem and discussion}

There are many equivalent definitions of infinitesimal local higher
symmetries for partial differential equations (see for instance
\cite{mr93b:58070}). In this paper we adopt the following
definition. A partial differential equation
$u_{\tau}=G(u_m,u_{m-1},\ldots,u,u_{s,t},u_{s-1,t},\ldots,u_t)$
generates a local symmetry of equation (\ref{utt}) if it is
compatible with equation (\ref{utt}) \cite{ssr}. A symmetry is
called a higher symmetry of equation (\ref{utt}) if $m>1$ or $s>1$.

Surprisingly enough equation (\ref{utt}) and its symmetry
(\ref{sym}) can be rewritten in a very compact form
\begin{equation}
\label{zt} z_t=z_n-z^2,
\end{equation}
\begin{equation}
\label{utau} u_{\tau}=z,
\end{equation}
where
\begin{equation}
\label{z} z=\frac{1}{u}(u_n-u_t)
\end{equation}
For any $n\ge 2$ equation (\ref{zt}) does not posses higher
symmetries and is not integrable by the inverse scattering method or
solvable by any other methods. Its trivial solution $z=0$
corresponds to a stationary point of  symmetry (\ref{utau}) and
provides a nontrivial family of solutions to equation (\ref{utt}).
Indeed, $z=0$ implies
\[
u_t=u_n.
\]
The general solution of this linear equation depends on one
arbitrary function and it can be found by standard methods.

\subsection*{Proof of the Theorem}
Let us rewrite equation (\ref{utt}) as a system of two evolutionary
equations on variables $u$ and $v=u_t$ as follows:
\begin{eqnarray}
\label{sys} u_t&=&v,\\ \nonumber
v_t&=&v_{n}-\frac{v}{u}(u_n-v)+\frac{(u_n-v)^2}{u}-u\partial_x^n\left(\frac{u_n-v}{u}\right)\equiv
H[u,v].
\end{eqnarray}
Equation (\ref{sym}) is a symmetry of (\ref{utt}) if and only if
system (\ref{sys}) is compatible with
\begin{eqnarray}
\label{symsys} u_{\tau}&=&\frac{1}{u}(u_n-v),\\ \nonumber
v_{\tau}&=&D_t\left(\frac{1}{u}(u_n-v)\right)=\partial_x^n\left(\frac{u_n-v}{u}\right)-\frac{(u_n-v)^2}{u^2},
\end{eqnarray}
where $D_t$ stands for the operator of total differentiation with
respect to $t$ according to (\ref{sys}). Calculating the cross
derivatives and expressing the results in terms of variable $z$
(\ref{z}) we obtain:
\begin{eqnarray}
\nonumber
&&u_{t\tau}=v_{\tau}=z_n-z^2,\\
\nonumber &&u_{\tau t}=z_t=z_n-z^2
\end{eqnarray}
and hence $u_{t\tau}=u_{\tau t}$. Analogously
\begin{eqnarray}
\nonumber
&&v_{t\tau}=\partial_x^n(z_n-z^2)-2z(z_n-z^2)-u\partial_{\tau}(z_n-z^2)-vz_{\tau},\\
\nonumber &&v_{\tau t}=\partial_x^n(z_n-z^2)-2z(z_n-z^2).
\end{eqnarray}
It is easy to verify that $z_{\tau}=0$ and therefore
$v_{t\tau}=v_{\tau t}$ and hence (\ref{sym}) is the symmetry of
equation (\ref{utt}).

Let us prove that symmetry (\ref{sym}) is the only local higher
symmetry of equation (\ref{utt}). In our proof we shall use elements
of the Perturbative Symmetry Approach in the symbolic representation
\cite{mr99g:35058,mn1}.

Function $H[u,v]$ in the right hand side of the system (\ref{sys})
can be rewritten in the form
\begin{equation}
\label{H} H[u,v]=-u_{2n}+2u_{n,t}+\sum_{i=1}^nf_i(u)P_i[u,v],
\end{equation}
where $f_i(u)=\frac{1}{u^i}$ and $P_i[u,v]$ are polynomials in $v$
and derivatives of $u$ and $v$ with respect to $x$ with constant
coefficients, such that
\begin{equation}
\label{Pord} P_i[\lambda u,\lambda
v]=\lambda^{i+1}P_i[u,v],\,\lambda\in C.
\end{equation}
Equation (\ref{utt}) is a homogeneous equation, therefore
polynomials $P_i$ are homogeneous polynomials of the weight $2n$,
i.e. $W(P_i)=2nP_i$, where
\[
W=\sum_{k=1}^{\infty}ku_k\frac{\partial}{\partial
u_k}+\sum_{k=0}^{\infty}(k+n)v_k\frac{\partial}{\partial v_k}.
\]
Function $H[u,v]$ can be treated as a differential polynomial in $v$
and derivatives of $u,v$ with coefficients being functions of $u$
only.  The symbolic representation of such polynomials have been
defined and studied in \cite{sw2}

To prove the second statement of the theorem we will need only a few
first terms of the function $H[u,v]$ (\ref{H}):
\begin{eqnarray}
\nonumber &&H[u,v]=-u_{2n}+2v_n+\\
\nonumber
&&+f_1(u)\left(\partial_x^n(uu_n-uv)-uu_{2n}+u_n^2+uv_n-3u_nv+2v^2\right)+R[u,v],
\end{eqnarray}
where $R[u,v]=\sum_{i\ge 2}f_i(u)P_i[u,v]$. Then in the symbolic
representation system (\ref{sys}) can be rewritten as
\begin{eqnarray}
\label{syssym} u_t&=&\hat{v},\\ \nonumber v_t&=&-\hat{u}
k_1^{2n}+2\hat{v} q_1^n+f_1(u)\circ
\left[\hat{u}^2a_1(k_1,k_2)+\hat{u}\hat{v}
a_2(k_1,q_1)+2\hat{v}^2)\right]+\hat{R}[u,v],
\end{eqnarray}
where
\begin{eqnarray}
\nonumber
a_1(k_1,k_2)&=&\frac{1}{2}\left[(k_1+k_2)^n(k_1^n+k_2^n)-(k_1^n-k_2^n)^2
\right],\\ \nonumber a_2(k_1,q_1)&=&q_1^n-(k_1+q_1)^n-3k_1^n,
\end{eqnarray}
and $\hat{R}[u,v]$ stands for the symbolic representation of
$R[u,v]$.

It is easy to verify that the most general form of a higher symmetry
($m\ge 2$) of system (\ref{sys}) in the symbolic representation is:
\begin{eqnarray}
\label{symsym} u_{\tau}&=&g_1(u)\circ \hat{u} k_1^m+g_2(u)\circ
\hat{v}
q_1^{m-n}+\hat{u}^2A_1(k_1,k_2;u)+\\ \nonumber &&+\hat{u}\hat{v}A_2(k_1,q_1;u)+\hat{v}^2A_3(q_1,q_2;u)+
\hat{S}[u,v]\equiv G,\\
\nonumber v_{\tau}&=&D_t(G).
\end{eqnarray}
Here $A_i(x,y;u)$ are polynomials in $x,y$ with coefficients,
depending on $u$. Polynomials $A_i(x,y;u)$ satisfy the condition
$A_i(0,y)=0,\,i=1,2$. The remainder $\hat{S}[u,v]$ stands for the
terms of higher nonlinearity in $v$ and $x$-derivatives of $u,v$.

\begin{Lemma} If (\ref{symsym}) is a symmetry of equation
(\ref{syssym}) then $g_1(u),g_2(u)$ are not equal to zero
simultaneously.
\end{Lemma}

{\it Sketch of the proof:} Let us assume that symmetry
(\ref{symsym}) does not have linear terms (in $\hat{u},\hat{v}$) and
starts with the terms of order $s>1$:
\[
u_{\tau}=\sum_{p=0}^sA_p(k_1,k_2,\ldots,k_p,q_1,q_2,\ldots,q_{s-p};u)\hat{u}^p\hat{v}^{s-p}+\mbox{higher
order terms}.
\]
Then it follows from the compatibility conditions of (\ref{syssym})
and (\ref{symsym}) that the coefficients
$A_p(k_1,k_2,\ldots,k_p,q_1,q_2,\ldots,q_{s-p};u)$ satisfy a linear
system of homogeneous algebraic  equations. This system has only a
trivial solution $A_p=0,\,\,p=0,\ldots,s$ since the determinant of
the corresponding matrix is not identically vanishing. Lemma is
proved $\blacksquare$.

The case $m=n$ corresponds to symmetry (\ref{sym}). In this case
$A_i(x,y;u)=0$ and $g_1(u)=-g_2(u)=\frac{1}{u}$.

Let $m\ne n$. From the compatibility conditions of (\ref{syssym})
and (\ref{symsym}) it follows that
\begin{eqnarray}
\nonumber
A_3(q_1,q_2;u)&=&\frac{1}{2}f_1(u)g_2(u)\frac{q_1^{m-n}+q_2^{m-n}-2(q_1+q_2)^{m-n}}{q_1^n+q_2^n-(q_1+q_2)^n}.
\end{eqnarray}
Polynomial $q_1^n+q_2^n-(q_1+q_2)^n$ does not divide
$q_1^{m-n}+q_2^{m-n}-2(q_1+q_2)^{m-n}$ if $m\ne n$ and therefore
$A_3(q_1,q_2;u)$ does not represent a symbol of a differential
polynomial with $u$-depended coefficients. It implies that
$g_2(u)=0$.

Having $g_2(u)=0$ we find (from the compatibility conditions) that
\begin{eqnarray}
\nonumber
A_2(k_1,q_1;u)&=&f_1(u)g_1(u)\frac{k_1^m+q_1^m-(k_1+q_1)^m}{k_1^n+q_1^n-(k_1+q_1)^n}.
\end{eqnarray}
For $s>1$ the polynomial $K_s(x,y)=x^s+y^s-(x+y)^s$ can be
factorized as
\begin{equation}
\label{K}
K_s(x,y)=xyT_s(x,y),\quad T_s(0,y)\ne 0
\end{equation}
Hence
\begin{equation}
\label{A2}
A_2(k_1,q_1;u)=f_1(u)g_1(u)\frac{T_m(k_1,q_1)}{T_n(k_1,q_1)}.
\end{equation}
It follows from (\ref{K}) $A_2(0,q_1;u)=0$ only if $g_1(u)=0$. It
follows from the Lemma, that a symmetry without a linear part does
not exist. The theorem is proved $\blacksquare$.

\noindent {\bf Acknowledgements.}  VSN is funded by a Royal Society
NATO/Chevening fellowship on the project {\it Symmetries and
coherent structures in non-evolutionary partial differential
equations}. AVM and VSN thank the RFBR Grant No. 02-01-00431 for
partial support.

\end{document}